\documentclass[runningheads,10pt]{llncs}

\usepackage{graphicx}

\usepackage[utf8]{inputenc} %
\usepackage[T1]{fontenc}    %
\usepackage{hyperref}       %
\usepackage{url}            %
\usepackage{booktabs}       %
\usepackage{amsfonts}       %
\usepackage{nicefrac}       %
\usepackage{microtype}      %

\usepackage{tikz}

\begin{document}
\title{Unsupervised Anomaly Detection for X-Ray Images}

\author{
Diana Davletshina\inst{1}\thanks{The first four authors have equal contribution.}, %
Valentyn Melnychuk\inst{1}, %
Viet Tran\inst{1}, %
Hitansh Singla\inst{1}, \\ %
Max Berrendorf\inst{2}, %
Evgeniy Faerman\inst{2}, %
Michael Fromm\inst{2}, %
Matthias Schubert\inst{2}
}
\authorrunning{Davletshina et al.}

\institute{%
Ludwig-Maximilians-Universität München, Munich, Germany %
\email{\{d.davletshina,v.melnychuk,viet.tran,hitansh.singla\}@campus.lmu.de} %
\and %
Lehrstuhl für Datenbanksysteme und Data Mining, %
Ludwig-Maximilians-Universität München, Munich, Germany %
\email{\{berrendorf,faerman,fromm,schubert\}@dbs.ifi.lmu.de} %
}

\maketitle              %

\begin{abstract}
Obtaining labels for medical (image) data requires scarce
and expensive experts. Moreover, due to ambiguous symptoms, single
images rarely suffice to correctly diagnose a medical condition. Instead,
it often requires to take additional background information such as the
patient’s medical history or test results into account. Hence, instead of
focusing on uninterpretable black-box systems delivering an uncertain
final diagnosis in an end-to-end-fashion, we investigate how unsupervised
methods trained on images without anomalies can be used to assist doctors
in evaluating X-ray images of hands. Our method increases the efficiency
of making a diagnosis and reduces the risk of missing important regions.
Therefore, we adopt state-of-the-art approaches for unsupervised learning
to detect anomalies and show how the outputs of these methods can
be explained. To reduce the effect of noise, which often
can be mistaken for an anomaly, we introduce a powerful preprocessing
pipeline. We provide an extensive evaluation of different approaches and
demonstrate empirically that even without labels it is possible to achieve
satisfying results on a real-world dataset of X-ray images of hands. We also
evaluate the importance of preprocessing and one of our main findings is
that without it, most of our approaches perform not better than random.
To foster reproducibility and accelerate research we make our code publicly available on GitHub\footnote{\url{https://github.com/Valentyn1997/xray}}. \end{abstract}

\section{Introduction}
Deep Learning techniques are ubiquitous and achieving state-of-the-art performance in many areas.
However, they require vast amounts of labeled data as witnessed by the marvelous boost in image recognition after the publication of the large scale ImageNet data set \cite{Deng2009,Krizhevsky2012}.
In medical applications, labels are expensive to acquire. While anyone can decide whether an image depicts a dog or a cat, deciding whether a medical image shows abnormalities, is a highly difficult task requiring specialists with years of training.
Another specialty of medical applications is that a simple classification decision does often not suffice.
End-to-end deep learning solutions tend to be hard to interpret, preventing their application in an area as sensitive as deciding for a treatment.
Moreover, additional patient information such as the patient's medical history, and clinical test results are often crucial to a correct diagnosis.
Integrating this information into an end-to-end pipeline is difficult and makes results even less interpretable.
Thus, the motivation for our work is to let doctors decide about the final diagnosis and treatment and develop a system, which can provide hints for doctors where to pay more attention to.

Hence, in this work, we investigate how we can support doctors to faster assess X-ray images, and reduce the chance of overlooking suspicious regions.
To this end, we demonstrate how state-of-the-art unsupervised methods, such as Autoencoders (AE) or Generative Adversarial Networks (GANs), can be used for anomaly detection on X-ray images.
As this dataset is noisy, and this is a general problem for a lot of real-world datasets, we present a sophisticated preprocessing pipeline to obtain %
better training data.
Afterwards, we train several unsupervised models, and explain for each, how to obtain several image-level anomaly scores.
For some of them, it is even natural to obtain pixel-wise annotations, highlighting anomalous regions.
One of our main findings is that accurate data preprocessing is indispensable.
The advantage of using autoencoders is that they naturally can provide pixel-level anomaly heatmaps, which can be used to understand model decisions.
In contrast, GAN-based approaches seem to be able to cope with more noisy data, yet being only able to produce image-wise anomaly scores.
We envision that this methodology can be easily installed in clinical daily routine to support doctors in quickly assessing X-ray images and spotting candidate regions for anomalies.

\begin{figure}
    \centering
    \includegraphics[width=.3\linewidth]{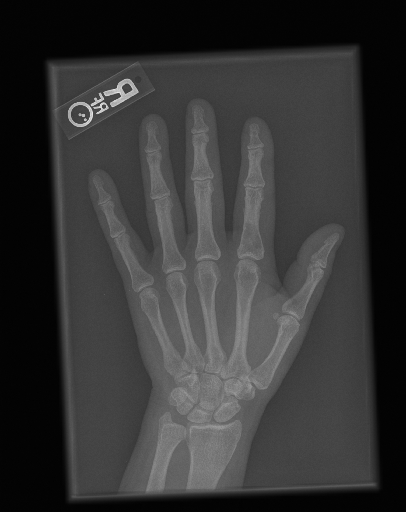}
    \includegraphics[width=.3\linewidth]{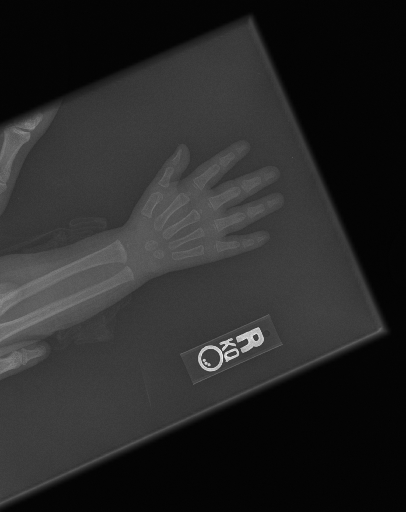}
    \includegraphics[width=.3\linewidth]{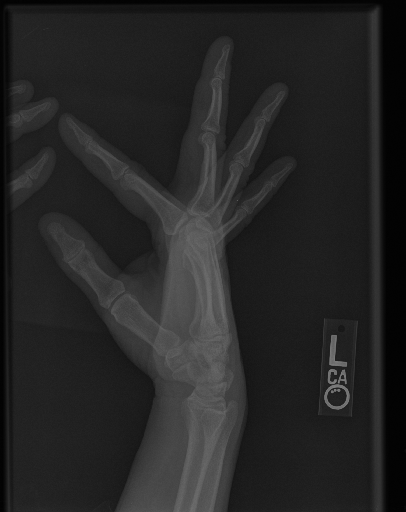}\\
    \includegraphics[width=.3\linewidth]{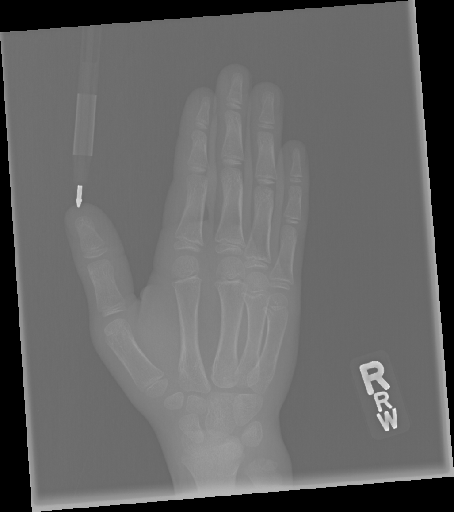}
    \includegraphics[width=.3\linewidth]{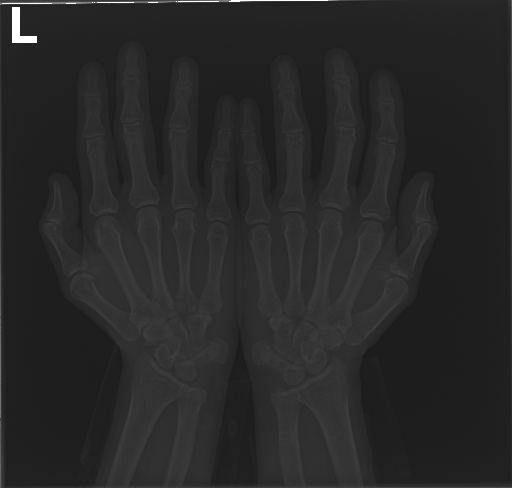}
    \includegraphics[width=.3\linewidth]{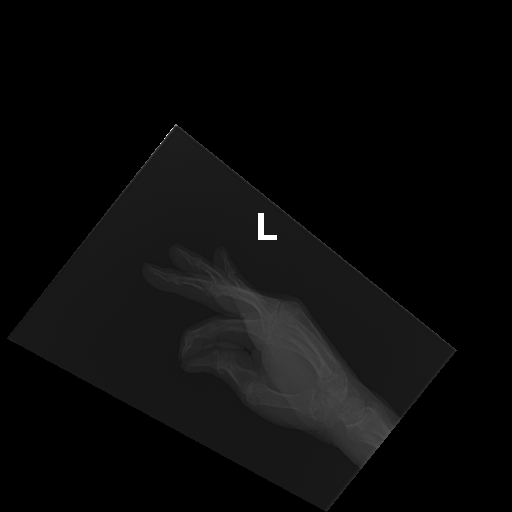}\\
    \includegraphics[width=.3\linewidth]{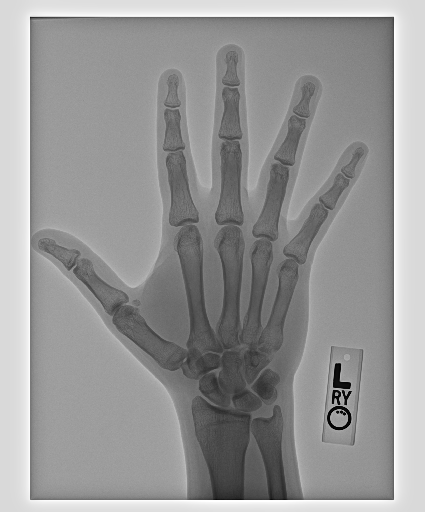}
    \includegraphics[width=.3\linewidth]{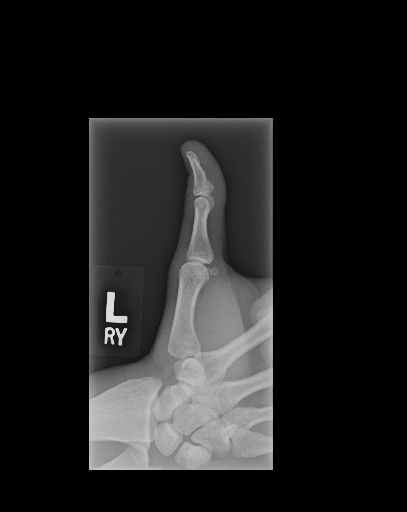}
    \includegraphics[width=.3\linewidth]{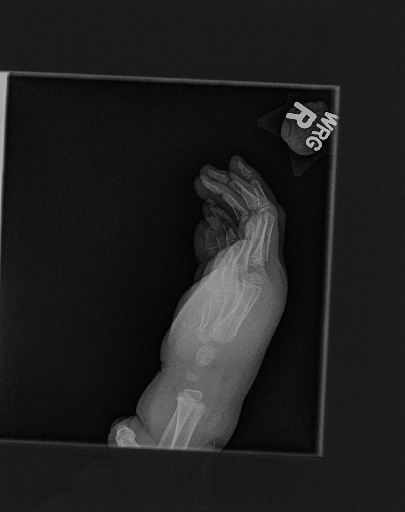}
    \caption{A few examples from the used subset of the MURA dataset containing X-ray images of hands demonstrating the large variety of image quality.}
    \label{fig:examples}
\end{figure}

In this work, we focus on a subset of the MURA dataset~\cite{Rajpurkar2017} containing only hand images.
In total, we have 5,543 images of 2,018 studies of 1,945 patients.
Each study is labeled as negative or positive, where positive means that there was an anomaly diagnosed in this study.
There are 521 positive studies, with a total of 1,484 images.
Figure~\ref{fig:examples} shows some examples from the dataset.
In summary, our contributions are as follows:
\begin{enumerate}
    \item We present a powerful preprocessing pipeline for the MURA dataset~\cite{Rajpurkar2017}, enabling the construction of a high-quality training set.
    \item We extensively survey unsupervised Deep Learning methods, and present approaches on how to obtain image-level and even pixel-level anomaly scores.
    \item We show extensive experiments on a real-world dataset evaluating the influence of proper preprocessing as well as the usability of the anomaly scores.
    To foster reproducibility, we will make our code public in the camera-ready version. 
\end{enumerate}
The rest of the paper is structured as follows:
In Section~\ref{sec:dataset} we describe our approach. We start with the description of data preprocessing in \ref{subsec:preprocessing} and describe anomaly detection approaches along with anomaly scores in section~\ref{subsec:models}. %
We discuss related work in section~\ref{sec:related_work}.
Finally, Section~\ref{sec:experiments} shows quantitative and qualitative experimental results on image-level and pixel-level anomaly detection.

\section{Unsupervised Anomaly Detection}
\label{sec:dataset}

\subsection{Preprocessing}
\label{subsec:preprocessing}
\begin{figure}
    \centering
    \scalebox{0.6}{
    \begin{tikzpicture}[
        every node/.style={draw, rectangle, minimum width=2.5cm, minimum height=1.5cm, align=center}, 
        offline/.style={text=black, fill=green, fill opacity=.2, text opacity=1},
        online/.style={text=black, fill=orange, fill opacity=.2, text opacity=1},
        model/.style={text=black, fill=blue, fill opacity=.2, text opacity=1},
        node distance=4cm,
    ]
    \draw node (input) {Input};
    \draw node[offline, right of=input] (crop) {Cropping};
    \draw node[offline, right of=crop] (center) {Hand Center\\Localization};
    \draw node[offline, right of=center] (semseg) {Hand\\Segmentation};
    \draw node[online, below of=semseg, yshift=2cm] (aug) {Augmentation};
    \draw node[online, left of=aug] (pad) {Padding \&\\Centering};
    \draw node[online, left of=pad] (norm) {Min-Max\\Normalization};
    \draw node[model, left of=norm] (model) {Model};
    \draw node[minimum width=1.5cm, minimum height=.4cm, left of=model, yshift=.4cm, fill=green, fill opacity=.2] (loffline){};
    \draw node[minimum width=1.5cm, minimum height=.4cm, left of=model, yshift=-.4cm, fill=orange, fill opacity=.2] (lonline) {};
    \draw node[draw=none, right of=loffline, node distance=1.5cm] {offline};
    \draw node[draw=none, right of=lonline, node distance=1.5cm] {online};
    \draw[-latex] 
          (input)   edge (crop)  
          (crop)    edge (center)
          (center)  edge (semseg)
          (semseg)  edge (aug)
          (aug)     edge (pad)
          (pad)     edge (norm)
          (norm)    edge (model)
          ;
    \end{tikzpicture}
    }
    \caption{The full image preprocessing pipeline. Steps highlighted in green are performed once and the result is stored to disk. Steps highlighted in orange are done on-the-fly.}
    \label{fig:pipeline}
\end{figure}
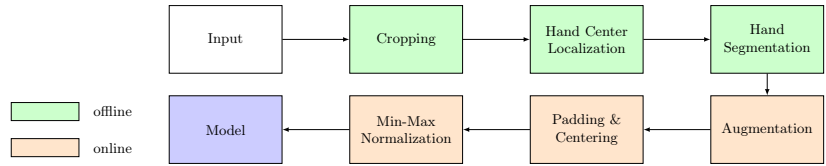

Real-life data is often noisy. This is especially problematic for unsupervised approaches for anomaly detection. On the one hand, it is necessary to remove noise to make sure that it is not recognized as an anomaly. On the other hand, it is crucial that the data denoising process does not mistake anomalies for noise and does not remove them. 
After experimenting a lot, we end up with the preprocessing pipeline 
depicted in Figure~\ref{fig:pipeline}.
We distinguish offline and online processing steps, where the offline processing is done once and then stored to disk to save time, whereas the online preprocessing is done on-the-fly while loading the data.
The individual steps are described in detail subsequently.

\begin{figure}
    \centering
    \includegraphics[height=3.5cm]{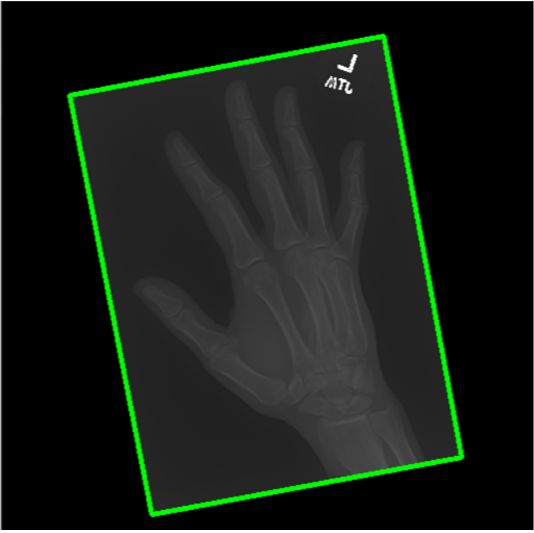}
    \includegraphics[height=3.5cm]{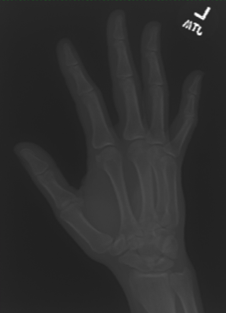}
    \includegraphics[height=3.5cm]{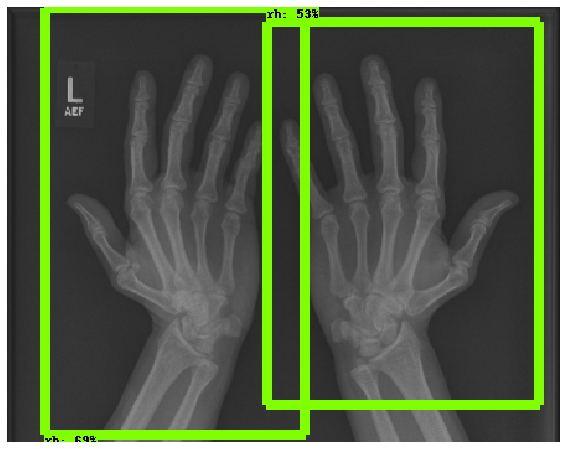}
    \caption{
    Result of image carrier detection with OpenCV (left side).
    The first image shows the original image with a detected rectangle.
    Next to it is the extracted image.
    The right image shows the result of running object detection on an image containing two hands. 
    We extract the image of both hands separately such that our preprocessed data set does not contain images with more than one hand.
    }
    \label{fig:crop}
    \label{fig:object_detection}
\end{figure}

\paragraph{Cropping}
The first step in our pipeline is to detect the X-ray image carrier in the image.
To this end, we apply OpenCV's contour detection using Otsu binarization \cite{Otsu1979}, and retrieve the minimum size bounding box, which does not need to be axis-aligned.
This works sufficiently well as long as the majority of the image carrier is within the image (cf. Figure~\ref{fig:crop}).
However, the approach might fail for heavily tilted images or those where larger parts of the image carrier reach beyond the image border.
\paragraph{Hand Localization} To further improve the detection of hands, and in particular split the images where two hands are depicted on one image, we manually labeled approximately 150 bounding boxes in the images.
Using this small dataset, we fine-tune a pre-trained single shot multibox detector (SSD) \cite{Liu2016} with MobileNet as taken from TensorFlow.
An exemplary results can be seen in Figure~\ref{fig:object_detection}.

\paragraph{Foreground Segmentation}
In a final step, foreground segmentation is performed using Photoshop's "select subject" method in batch processing mode.
Thereby, we obtain a pixel-wise mask, roughly encompassing the scanned hand.

\paragraph{Data Augmentation}
Due to GPU memory constraints, the images for BiGAN and $\alpha$-GAN are resized to 128 pixels on the longer image side while maintaining aspect ratio before applying the augmentation.
For the auto-encoder models, this is not necessary.
Afterwards, standard data augmentation methods (horizontal/vertical flipping, channel-wise multiplication, rotation, scaling) using the \texttt{imgaug}\footnote{https://github.com/aleju/imgaug} library are applied before finally padding the images to 512x512 (AE + DCGAN) or 128x128 (BiGAN + $\alpha$GAN) pixels.

\subsection{Models}
\label{subsec:models}
In this section, we describe the different model types we trained in a fully unsupervised / self-supervised fashion on the train data part comprising only images from patients without attested anomalies.
We also describe how to obtain anomaly scores from the trained models.
In the appendix %
we additionally provide details about the architecture for every model.

\subsection{Autoencoders}
We studied different auto-encoder architectures for the task at hand.
Common among them is their usage of a reconstruction loss, i.e. the input to the network is also used as the target, and we evaluate how well the input is reconstructed.
As the information has to pass through an informational bottleneck, the model cannot simply copy the input data, but instead has to perform a form of compression, extracting features which suffice to reconstruct the image sufficiently well.
Hence, we have an encoder part of the network ($E$), which transforms the input $\mathbf{x}$ non-linearly to a latent space representation $\mathbf{z}$.
Analogously, there is a decoder $D$ that transforms an input $\mathbf{z}$ from latent space back to and element $\hat{\mathbf{x}}$ in the input space.

For simplicity, we describe the general loss formulation using a vector input $\mathbf{x} \in \mathbb{R}^n$ instead of a two-dimensional pixel-matrix.
In its simplest form, the reconstruction loss is given as the mean over pixel-wise squared differences.
Let $\mathbf{x}, \hat{\mathbf{x}} = D(E(\mathbf{x})) \in \mathbb{R}^n$, then
$$
\mathcal{L}(\mathbf{x}, \hat{\mathbf{x}}) = \frac{1}{n} \sum \limits_{i=1}^n (x_i - \hat{x}_i)^2
= \frac{1}{n}\|\mathbf{x} - \hat{\mathbf{x}}\|_2^2
$$

As we are only interested in detecting anomalies on the hand part of the image, we consider a variant of this loss, named \emph{masked reconstruction loss}, where only those pixels are considered that belong to the mask.
Let $\mathbf{m} \in \{0, 1\}^n$ be the mask, where $m_i = 1$ if and only if the position $i$ belongs to the hand.
Then,
$$
\mathcal{L}_M(\mathbf{x}, \hat{\mathbf{x}}, \mathbf{m}) = \frac{1}{\|\mathbf{m}\|_1} \|\mathbf{m} \odot (\mathbf{x} - \hat{\mathbf{x}})\|_2
$$
where $\odot$ denotes the Hadamard product (i.e. element-wise multiplication).
In the following, we describe the architectures of the network in more detail.

\paragraph{}
In a \emph{convolutional auto-encoder (CAE)}, we implement encoder and decoder as fully convolutional neural networks (CNNs).
In general, the encoder is built as a sequence of repeated convolution blocks.
We apply Batch Normalization \cite{Ioffe2015} between every convolution and the respective activation, and use ReLU \cite{Glorot2011} as activation function.
A detailed model description is given in the appendix.
Similarly, the decoder consists of repeated blocks of transposed convolutions.
As before, we apply batch normalization before every activation function.
As bottleneck size, we use a spatial resolution of $16 \times 16$ and 512 channels.

\paragraph{}
\emph{Variational AE (VAE)} \cite{Kingma2013} is a generative model, which maps an input to a Gaussian distribution in latent space, characterized by its mean and covariance $(\mathbf{\mu(x)}, \mathbf{\Sigma(x)})$, instead of mapping it to a fixed latent representation.
The covariance matrix is usually restricted to a diagonal matrix.
For reconstruction, a sample $\mathbf{z} \sim \mathcal{N}(\mathbf{\mu(x)}, \mathbf{\Sigma(x)})$ is drawn, and passed through the decoder sub-network.
To avoid very small values in $\mathbf{\Sigma}$, and thereby approaching a delta distribution, i.e. traditional AE, an additional loss term is introduced as the Kullback-Leibler divergence (KLD) between $\mathcal{N}(\mathbf{\mu(x)}, \mathbf{\Sigma(x)})$ and the standard normal distribution $\mathcal{N}(\mathbf{0}, \mathbf{I})$.

\subsubsection{Anomaly Detection Scores}
The rationale behind using AE for anomaly detection is that as the AE is trained on normal data only, it has not seen anomalies during training, and hence will fail to reproduce them.
Due to the convolutional nature of the network, the error is even expected to occur stronger in regions close to the anomaly, and less strong further apart.
If the receptive field is small enough, those regions outside of it are not affected at all.
Hence, we can use the reconstruction error in two ways:
\begin{enumerate}
    \item Pixel-wise to obtain a heatmap highlighting regions that were hardest to reconstruct.
    If there is an anomaly, we expect the highest error in that region.
    We show an example for such in the qualitative results, Figure~\ref{fig:heatmap}.
    \item Aggregated over all pixels (under the mask) to obtain an image-wise score.
    As for aggregation, we explore different aggregation strategies. In the simplest case, we just average over all locations.
    By using only the highest $k$ values to compute the mean, we can obtain a score that is more sensitive towards regions of high reconstruction error (i.e. anomalous regions).
\end{enumerate}
We aim for using auto-encoder architectures which are strong enough to successfully reconstruct normal hands, without risking to learn identity mappings by allowing too wide bottlenecks.
While the architecture should generalize over all normal hands, a too strong generalization might cause the effect that also anomalies can be reconstructed sufficiently well.

\subsection{GAN}
A Generative Adversarial Network (GAN) \cite{Goodfellow2014} comprises two sub-networks, a generator $G$, and a discriminator $D$, which can be seen as antagonists in a two-player game.
The generator takes random noise as input and generates samples in the target domain.
The discriminator takes real data points, as well as generated ones, and has to distinguish between real and fake data.
The sub-networks are trained alternatingly, and if successful, the generator can afterwards be used to sample from the (approximated) data distribution, and the discriminator can be used to decide whether a sample is drawn from the given data distribution.

\paragraph{}
\emph{Deep Convolutional GAN (DCGAN)}\cite{Radford2015} is an extension of the original GAN architecture to convolutional neural networks.
Similarly to the CAE, the two networks contain convolutions (discriminator) and transposed convolutions (generator) instead of the fully connected layers of the originally proposed GAN architecture.

\paragraph{}
\emph{BiGAN \cite{Donahue2017} / ALI \cite{Dumoulin2017}} extends DCGAN by an encoder $E$, which encodes the real image into latent space\footnote{In the paper $E$ is called $G_z(\mathbf{x})$ as opposed to the generator $G = G_x(\mathbf{z})$}.
The discriminator is now provided with both, the real and fake image, as well as their latent codes, i.e. $D((G(\mathbf{z}), \mathbf{z}), (\mathbf{x}, E(\mathbf{x})))$.

\paragraph{}
\emph{$\alpha$-GAN} \cite{Rosca2017} comprises four sub-networks:
\begin{itemize}
    \item An encoder $E(\mathbf{x})$ which transforms a real image into a latent representation.
    \item A code-discriminator $CO(\mathbf{z})$ which distinguishes between the latent representations produced by the encoder and random noise used as generator input.
    \item A generator $G(\mathbf{z})$ which generates an image from either the randomly sampled $\mathbf{z}$, or the encoded image $E(\mathbf{x})$.
    \item A discriminator $D(\mathbf{x})$ which distinguishes between reconstructed real images $G(E(\mathbf{x}))$, and generated images $G(\mathbf{z})$.
\end{itemize}
In addition to the classification losses for both discriminators, a reconstruction loss is applied for the auto-encoder formed by the encoder-generator pair.
Hence, the code-discriminator gives the encoder the incentive to transform the inputs to match the random distribution, similarly as in VAE through the KL-divergence.
Likewise, the discriminator motivates matching the data distribution in the image domain.

\subsubsection{Anomaly Detection Scores}
For the GAN models, we generally use the discriminator's output as the anomaly score.
When converged, the discriminator should be able to distinguish between images belonging to the data manifold, i.e. images of hands without any anomalies, and those which lie outside, such as those containing anomalous regions.
For $\alpha$GAN we use the mean over code discriminator and discriminator probability.

\section{Related Work}
\label{sec:related_work}
With the rapid advancement of deep learning methods, they have also found their way into medical imaging, cf. e.g. \cite{Litjens2017,Raza2018}.
Despite the limited availability of labels in medical contexts, supervised methods make up the vast majority.
Very likely, this is due to the easier trainability, but possibly also because the interpretability of the results so far has often been secondary.
Sato et al.~\cite{Sato2018} use a 3D CAE for a pathology detection method in CT scans of brains.
The CAE is trained solely on normal images, and at test time, the MSE between the image and its reconstruction is taken as the anomaly score.
Uzunova et al.~\cite{Uzunova2019} use VAE for medical 2D and 3D CT images. 
Similarly, they use MSE reconstruction loss as the anomaly score.
Besides the KL-divergence in latent space, they use a $L_1$ reconstruction loss for training, which produced less smooth output. GANomaly \cite{Samet2019} and its extension with skip-connections uses an AE and maps the reconstructed input back to the latent space. The anomaly score is computed in latent space between original and reconstructed input. They apply their methods on X-Ray security imagery to detect anomalous items in baggage.
Recently, there have been a lot of publications using the currently popular GANs. 
For example, \cite{Madani2018} uses a semi-supervised approach for anomaly detection in chest X-ray images.
They replace the standard discriminator classification into real and fake, with a three-way classification into real-normal, real-abnormal, and fake.
While this allows training with fewer labels, it still requires them for training.
Schlegl et al \cite{Schlegl2017}, train a DC-GAN on slices of OCT scans, where the original volume is cut along the x-z axis, and the slices are further randomly cropped.
At test time, they use gradient descent to iteratively solve the inverse problem of obtaining a latent vector that produces the image.
Stopping after a few iterations, the $L_1$ distance between the generated image and the input image is considered as residual loss.
To summarize, the focus of recent work for anomaly detection approaches lies either in applying existing methods for a new type of data or adapting unsupervised methods for anomaly detection. Instead, we provide an extensive evaluation of state-of-the-art unsupervised learning approaches that can be directly used for anomaly detection. Furthermore, we evaluate the importance of different preprocessing steps and compare methods with regard to explainability.

\section{Experiments}
\label{sec:experiments}
We demonstrate the capability of our preprocessing pipeline and all described models in experiments on a subset of the MURA dataset containing only X-ray images of hands. 3,062 images are stored in a single-channel PNG image, and 2,481 are stored with three RGB channels.
However, all images look like gray-scale images, which is why we convert all 3-channel images to a single channel.
The longest side of the images is always 512 pixels in size.
The smaller side ranges from 160 to 512, with the majority between 350 and 450.
\begin{figure}
    \centering
    \resizebox{.8\linewidth}{!}{
    \begin{tikzpicture}[xscale=.45]
    \definecolor{train}{RGB}{255,255,204}
    \definecolor{val}{RGB}{120,198,121}
    \definecolor{test}{RGB}{0,104,55}
    \definecolor{normal}{RGB}{254,204,92}
    \definecolor{abnormal}{RGB}{240,59,32}
    \draw[black, fill=train, fill opacity=.5] (0, 0) rectangle +(9.03, 1);
    \draw[black, fill=val, fill opacity=.5] (9.03, 0) rectangle +(5.21, 1);
    \draw[black, fill=test, fill opacity=.5] (14.24, 0) rectangle +(5.21, 1);
    \draw[black, fill=normal, fill opacity=.5] (0, 0) rectangle +(9.03, -1);
    \draw[black, fill=abnormal, fill opacity=.5] (9.03, 0) rectangle +(2.605, -1);
    \draw[black, fill=normal, fill opacity=.5] (11.635, 0) rectangle +(2.605, -1);
    \draw[black, fill=abnormal, fill opacity=.5] (14.24, 0) rectangle +(2.605, -1);
    \draw[black, fill=normal, fill opacity=.5] (16.845, 0) rectangle +(2.605, -1);
    \draw (4.515, 0.5) node {train};
    \draw (11.635, 0.5) node {validation};
    \draw (16.845, 0.5) node {test};
    \draw (-1.5, 0.5) node {split};
    \draw (-1.7, -0.5) node {anomaly};
    \draw (4.515, -0.5) node {n};
    \draw (10.3325, -0.5) node {p};
    \draw (12.9375, -0.5) node {n};
    \draw (15.5425, -0.5) node {p};
    \draw (18.1475, -0.5) node {n};
    \draw (0, -1.2) edge[-latex] +(19.45, 0);
    \foreach \x/\la in {0/0,2.5/250,5/500,7.5/750,10/1000,12.5/1250,15/1500,17.5/1750}{
        \draw (\x, -1.3) -- +(0, .1);
        \draw (\x, -1.5) node {\small \la};
    }
    \draw (9.25, -1.9) node {\#patients};
    \end{tikzpicture}
    }
    \caption{
    Visualization of the applied data split scheme.
    "n" denotes patients which do not have a abnormal study ("negative"), "p" the contrary ("positive").
    Notice that the training part of the split does not contain any images of anomalies, i.e. we do not use anomalous images for training.
    }
    \label{fig:split}
\end{figure}
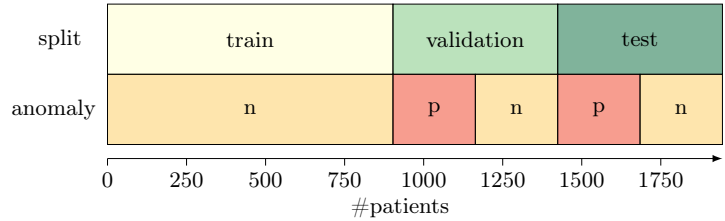

As our approach is unsupervised, we train only on negative images, i.e. images without an anomaly.
Furthermore, to avoid test leakage, we split the data by patient, and not by study or image, to ensure that we do not have an image of a patient in the training data, and another image of the same patient in the test or validation data.
To this end, we proceed as follows:
Let $P$ be the set of all patients, and $P^+$ be the set of patients with a study that is labeled as abnormal.
The rest of the patients is denoted by $P^- := P \setminus P^+$.
For the test and validation set, we aim at having balanced classes.
Therefore, we distribute $P^+$ evenly at random across test and validation.
Afterwards, we randomly sample the same number of patients without known anomalies for test and validation and use the rest of the patients for training.
The procedure is visualized in Figure~\ref{fig:split}.
In total, we end up with 2,554 training images, 1,494 validation images, and 1,495 test images.

We trained all models on a machine with one NVIDIA Tesla V100 GPU with 16GiB of VRAM, 20 cores and 360GB of RAM.
Following \cite{Raghu2019}, we train our models from scratch and do not use transfer learning from large image classification datasets.
We performed a manual hyper-parameter search on the validation set and selected the best-performing models per type with respect to Area-under-Curve for the Receiver-Operator-Curve (ROC-AUC).
We report the ROC-AUC on the test set. %

\subsection{Quantitative Results}
\begin{table}[h]
    \centering
    \caption{
    Quantitative results for all models. We report ROC-AUC on the test set for the best configuration regarding validation set ROC-AUC. All numbers are mean and standard deviation across four separate trainings with different random seeds. For each model we report results for various anomaly scores: Mean Squared Error (MSE), L1, Kullback–Leibler Divergence (KLD), Discriminator Probability (D). Top-200 denotes the case, when only 200 pixels with the highest error are taken into consideration.
    }
    \label{tab:quant}
    \scriptsize
    \tiny
    \begin{tabular*}{\linewidth}{l*{6}{@{\extracolsep{\fill}}c@{\extracolsep{\fill}}}}
    \toprule
    {} &                \multicolumn{2}{c}{\textbf{raw}} &               \multicolumn{2}{c}{\textbf{crop}} &               \multicolumn{2}{c}{\textbf{full}} \\
    {} & w/o HE & w/ HE & w/o HE & w/ HE & w/o HE & w/ HE \\
    \midrule
\textbf{CAE} & & & & & & \\
MSE &  .460  $\pm$  .033 &  .504  $\pm$  .034 &  .466  $\pm$  .022 &  .510  $\pm$  .021 &  .501  $\pm$  .013 &  \textbf{.570  $\pm$  .019} \\
MSE (top-200) &  .466  $\pm$  .013 &  .448  $\pm$  .025 &  .486  $\pm$  .015 &  .473  $\pm$  .018 &  .506  $\pm$  .039 &  .553  $\pm$  .023 \\
\midrule
\textbf{VAE}  & & & & & & \\
KLD &  .488  $\pm$  .031 &  .491  $\pm$  .013 &  .470  $\pm$  .046 &  .496  $\pm$  .045 &  .520  $\pm$  .026 &  \textbf{.533  $\pm$  .014} \\
L1 &  .432  $\pm$  .033 &  .446  $\pm$  .016 &  .438  $\pm$  .033 &  .438  $\pm$  .016 &  .435  $\pm$  .014 &  .483  $\pm$  .009 \\
L1 + KLD &  .432  $\pm$  .033 &  .446  $\pm$  .016 &  .438  $\pm$  .034 &  .437  $\pm$  .016 &  .438  $\pm$  .011 &  .488  $\pm$  .011 \\
L1 (top-200) &  .438  $\pm$  .017 &  .472  $\pm$  .010 &  .440  $\pm$  .025 &  .471  $\pm$  .013 &  .428  $\pm$  .013 &  .481  $\pm$  .010 \\
MSE &  .432  $\pm$  .033 &  .446  $\pm$  .016 &  .438  $\pm$  .033 &  .438  $\pm$  .016 &  .435  $\pm$  .014 &  .483  $\pm$  .009 \\
MSE + KLD &  .432  $\pm$  .033 &  .446  $\pm$  .016 &  .438  $\pm$  .033 &  .438  $\pm$  .016 &  .436  $\pm$  .013 &  .486  $\pm$  .010 \\
MSE (top-200) &  .438  $\pm$  .017 &  .472  $\pm$  .010 &  .440  $\pm$  .025 &  .471  $\pm$  .013 &  .428  $\pm$  .013 &  .481  $\pm$  .010 \\
\midrule
\textbf{DCGAN} & & & & & & \\
Disc. (D) &  .497  $\pm$  .018 &  .491  $\pm$  .041 &  .493  $\pm$  .015 &  .493  $\pm$  .025 & \textbf{.530  $\pm$  .027} &  .527  $\pm$  .022 \\
\midrule
\textbf{BiGAN}  & & & & & & \\
MSE &  .471  $\pm$  .021 &            - &  .438  $\pm$  .039 &            - &  .491  $\pm$  .042 &  .522  $\pm$  .017 \\
MSE (top-200) &  .471  $\pm$  .011 &            - &  .459  $\pm$  .030 &            - &  .475  $\pm$  .033 &  .508  $\pm$  .026 \\
Disc. (D) &  .508  $\pm$  .007 &            - &  .534  $\pm$  .016 &            - & \textbf{ .549  $\pm$  .006} &  .522  $\pm$  .019 \\
\midrule
\textbf{$\alpha$GAN}  & & & & & & \\
Code-Disc. (C) &  .500  $\pm$  .000 &            - &  .500  $\pm$  .001 &            - &  .500  $\pm$  .000 &  .500  $\pm$  .000 \\
MSE &  .476  $\pm$  .029 &            - &  .466  $\pm$  .022 &            - &  .442  $\pm$  .013 &  .528  $\pm$  .018 \\
MSE (top-200) &  .465  $\pm$  .031 &            - &  .446  $\pm$  .018 &            - &  .422  $\pm$  .016 &  .533  $\pm$  .013 \\
Disc. (D) &  .503  $\pm$  .022 &            - &  .534  $\pm$  .022 &            - &  \textbf{.607  $\pm$  .016} &  .584  $\pm$  .012 \\
C + D &  .503  $\pm$  .022 &            - &  .534  $\pm$  .022 &            - &  \textbf{.607  $\pm$  .016} &  .584  $\pm$  .012 \\
\bottomrule
\end{tabular*}
\end{table}

Apart from the performance of single models, we also evaluate the importance of the preprocessing steps.
Therefore, we evaluate the models on the raw data, the data after cropping the hand regions, as well as on the fully preprocessed data.
We also vary whether histogram equalization is applied before the augmentation or not.
We summarize the quantitative results in Table~\ref{tab:quant} showing the mean and standard deviation across four runs.
There is a clear trend in preprocessing:
All models have their best runs in the fully preprocessed setting, emphasizing the importance of our preprocessing pipeline for noisy datasets.
Interestingly, without foreground segmentation, i.e. only by cropping the single hands, the results appear to be worse than on the raw data.
While histogram equalization is a contrast enhancement method in particular useful to improve human perception of low-contrast images, it seems to improve the results for AE-based models consistently.
For BiGAN and $\alpha$GAN our experiments did not finish until the deadline.
As they comprise AE components we expect to see an improvement there.
On raw and also cropped data we frequently observe ROC-AUC values smaller than 45\%.
Hence, we might be able to improve the ROC-AUC score by flipping the anomaly decision.
Partially, we attribute this also to the rather unstable results for these models.
Regarding the aggregation of reconstruction error, we observe that using only the top-k loss values across all pixels does not improve the result.
We attribute that partially to not tuning enough across different values for $k$, as we only used $k=200$ for all models, which may be too few pixels to detect some anomalies.
Due to the lack of pixel-level annotation, we did not investigate this issue so far.
In total, we obtain the best ROC-AUC score with 60.7\% for $\alpha$-GAN using the discriminator probability.
CAE however also achieves 57\% ROC-AUC and additionally can naturally provide pixel-level anomaly scores yielding higher interpretability.

\subsection{Qualitative Results}
\begin{figure}
    \centering
    \includegraphics[width=.4\linewidth]{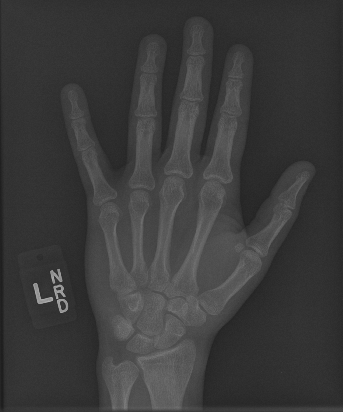}
    \includegraphics[width=.4\linewidth]{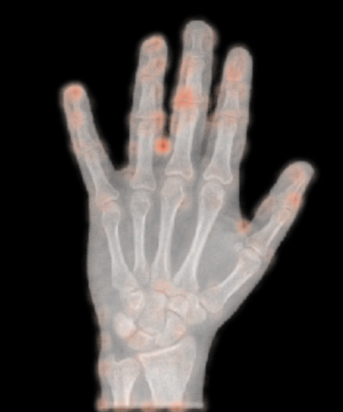}
    \\
    \includegraphics[width=.81\linewidth]{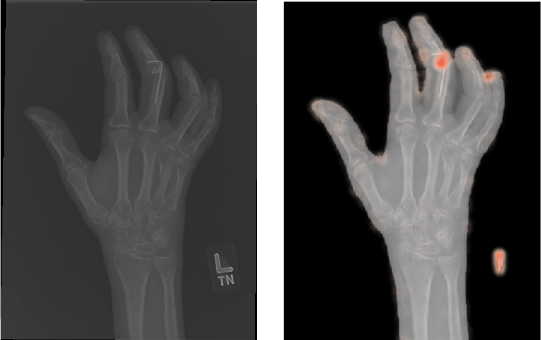}
    \caption{Example heatmaps of reconstruction error of CAE. The left image-pair shows a hand from a study labeled as normal hand. Here we can see that the reconstruction error is relatively wide spread. The right image pair shows an abnormal hand, where the abnormality is clearly highlighted.}
    \label{fig:heatmap}
\end{figure}
In addition to the numerical results we also showcase some qualitative results.
For all methods with reconstruction loss, i.e. all AE as well as $\alpha$-GAN, we can generate heatmaps visualizing the pixel-wise losses.
Thereby, we can highlight regions that could not be reconstructed well.
Following our assumption, these regions should be the anomalous regions.
In Figure~\ref{fig:heatmap}, we can see prototypical examples produced by CAE.
The upper image shows a hand contained in a study which was labeled as normal.
We can see that the reconstruction error does not occur concentrated, but is rather spread widely across the hand.
The maxima seem to occur around joints, which due to their more complex structure are likely to be harder to reconstruct.
Compared to the lower image, which shows a study labeled as abnormal, we see a clear highlighting at the middle finger.
Visible also for a non-expert, we can spot metal parts in the X-ray image at the very same location.
For those anomalies which could be validated by a person without a medical background, the highlighted regions seem to correspond largely to those anomalous regions.

\section{Conclusion}
In this paper, we investigated methods for unsupervised anomaly detection in X-ray images.
To this end, we surveyed two families of unsupervised models, auto-encoders and GANs, regarding their applicability to derive anomaly scores.
In addition, we provide a sophisticated multi-step preprocessing pipeline.
In our experiments, we compare the methods against each other, and furthermore, reveal that the preprocessing is crucial for most models to obtain good results on real-world data.
For the auto-encoder family, we study the interpretability of pixel-wise losses as anomaly heatmap and verify that in cases of anomalies which a non-expert can detect (e.g. metal pieces in the hand), these heatmaps closely match the anomalous regions.
As future work, we envision the extension to broader datasets such as the full MURA dataset, as well as obtaining pixel-level anomaly scores for the GAN based models.
To this end, methods from the field of explainable AI, such as grad-CAM \cite{Selvaraju2017} or LRP \cite{Bach2015} can be applied to the discriminator to obtain heatmaps similarly to those of the AE models.
Moreover, we see the potential for different model architectures closer tailored to the specific problem and data type, as well as the possibility of building an ensemble model using the different ways how to extract anomaly scores from single models, or even across different model types.

\section*{Acknowledgement}
We would like to thank Franz Pfister and Rami Eisaway from deepc (\url{www.deepc.ai}) for access to the data and support in understanding the use case.
Part of this work has been conducted during a practical course at Ludwig-Maximilians-Unversität München funded by Z.DB.
The infrastructure for the course was provided by the Leibniz-Rechenzentrum.
This work has been funded by the German Federal Ministry of Education and Research (BMBF) under Grant No. 01IS18036A. The authors of this work take full responsibilities for its content.
 
\bibliographystyle{abbrv}
\bibliography{main}

\clearpage
\appendix
\section*{Supplementary Material}
\subsection{Schematic Architectures and Reconstruction Examples}
\begin{figure}[h]
    \centering
    \includegraphics[width=\linewidth]{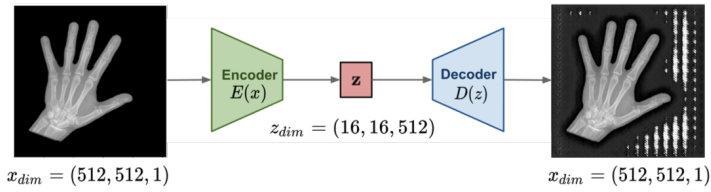}
    \caption{Schematic overview of convolutional autoencoder (CAE) and an example reconstruction of completely preprocessed image. Encoder $E(\mathbf{x})$ and decoder $D(\mathbf{z})$ are realized as
    deep convolutional neural networks (CNNs). Note, that masked reconstruction loss is used, therefore reconstruction outside of a hand is arbitrary.}
    \label{fig:cae}
\end{figure}

\begin{figure}[h]
    \centering
    \includegraphics[width=\linewidth]{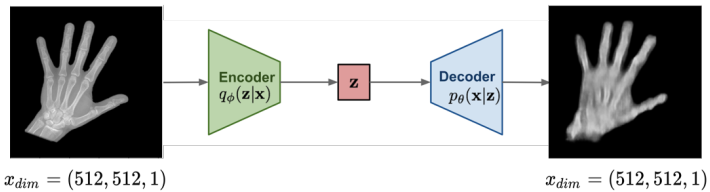}
    \caption{
    Schematic overview of variational autoencoder and an example reconstruction of completely preprocessed image.
    Encoder and decoder are realized as CNNs. 
    The encoder predicts a mean $\mu(\mathbf{x})$ and a covariance $\Sigma(\mathbf{x})$ of a Gaussian distribution in latent space.
    The reconstruction is done from a sample from this Gaussian.
    }
    \label{fig:vae}
\end{figure}

\begin{figure}[h]
    \centering
    \includegraphics[width=\linewidth]{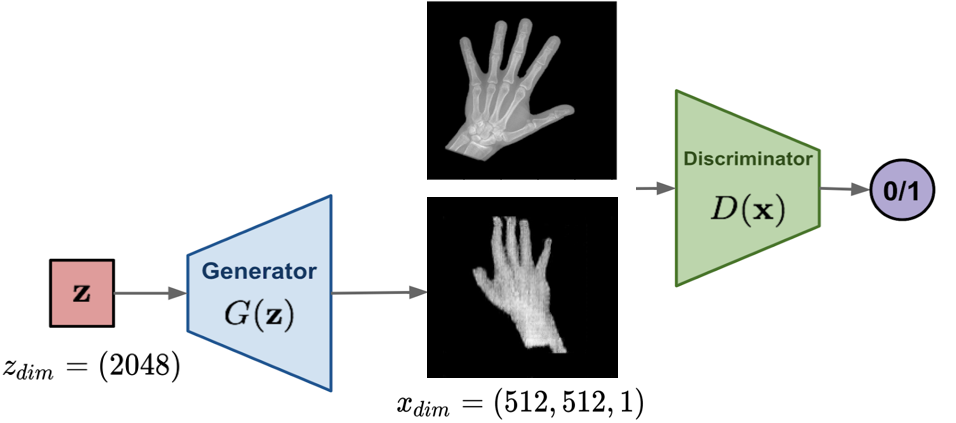}
    \caption{Schematic overview of DCGAN and example comparison of real (above) and fake (below) images. The generator $G(\mathbf{z})$ generates an image from input noise $\mathbf{z}$. The discriminator distinguishes between real images and generated ones. More details in the text.}
    \label{fig:dcgan}
\end{figure}

\begin{figure}[h]
    \centering
    \includegraphics[width=\linewidth]{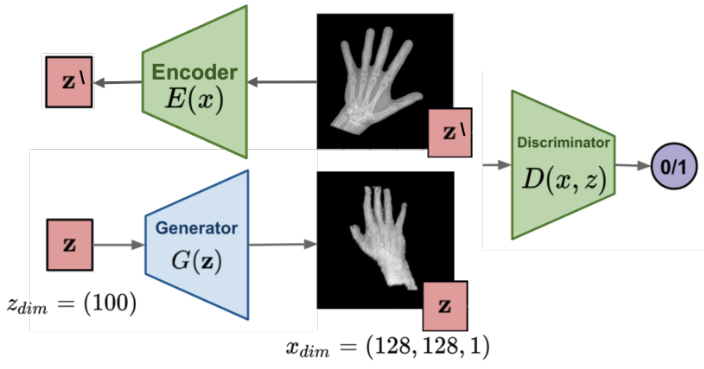}
    \caption{Schematic overview of BiGAN / ALI and an examples of real (above) and fake (below) images. The generator $G(\mathbf{z})$ generates an image from input noise $\mathbf{z}$. The encoder $E(\mathbf{x})$ encodes the real image. The discriminator distinguishes between real images and generated ones additionally provided with the noise vector $\mathbf{z}$ and the encoding $\mathbf{z}^{\backslash}$.}
    \label{fig:bigan}
\end{figure}

\begin{figure}[h]
    \centering
    \includegraphics[width=\linewidth]{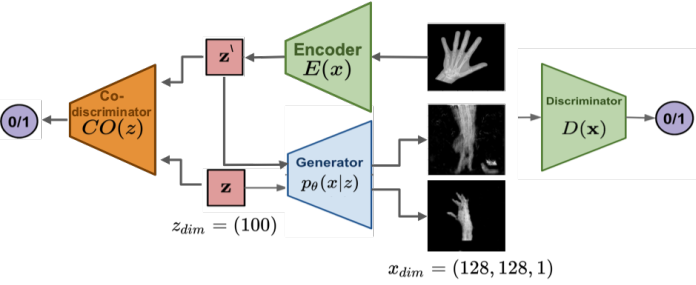}
    \caption{
    Schematic overview of $\alpha$-GAN comprising four sub-networks: encoder $E$, generator $G$, discriminator $D$ and code-discriminator $CO$.
    Discriminator and code-discriminator distinguish real from fake data in image space, and latent space, respectively.
    In addition, encoder and generator also form an auto-encoder.
    The image above is completely preprocessed real image.
    Its reconstruction is shown in the middle.
    The last image is fake produced by generator from random noise.
    }
    \label{fig:alphagan}
\end{figure}

\clearpage
\subsection{Architecture Details}
{
    \centering
    \resizebox{.3\linewidth}{!}{%
    \begin{tabular}[t]{ll}
        \multicolumn{2}{c}{CAE}\\
        \toprule
        \multicolumn{2}{c}{encoder}\\
        kernel size & output filters \\
        \midrule
         (3, 3) & (512, 512, 16) \\
         (4, 4) & (256, 256, 32) \\
         (3, 3) & (256, 256, 32) \\
         (4, 4) & (128, 128, 64) \\
         (3, 3) & (128, 128, 64) \\
         (4, 4) & (64, 64, 128) \\
         (3, 3) &  (64, 64, 128)\\
         (4, 4) & (32, 32, 256) \\
         (3, 3) &  (32, 32, 256)\\
         (4, 4) & (16, 16, 512) \\
        \midrule
        \multicolumn{2}{c}{decoder}\\
        kernel size & output filters \\
        \midrule
         (4, 4) & (32, 32, 256) \\
         (4, 4) & (64, 64, 128) \\
         (4, 4) & (128, 128, 64) \\
         (4, 4) & (256, 256, 32) \\
         (4, 4) & (512, 512, 16) \\
         (3, 3) & (512, 512, 1) \\
        \bottomrule
    \end{tabular}%
    }
    \resizebox{.3\linewidth}{!}{%
    \begin{tabular}[t]{ll}
        \multicolumn{2}{c}{VAE}\\
        \toprule
        \multicolumn{2}{c}{encoder}\\
        kernel size & output filters \\
        \midrule
        (4, 4) & (255, 255, 8) \\
        (4, 4) &  (126, 126, 16)\\
        (4, 4) & (62, 62, 32) \\
        (4, 4) & (30, 30, 64) \\
        (4, 4) &  (14, 14, 128)\\
        (4, 4) & (6, 6, 256) \\
        (4, 4) & (2, 2, 512) \\
        \midrule
        \multicolumn{2}{c}{bottleneck}\\
        \midrule
        reshape: $z$ & $(2 \cdot 2 \cdot 512)$\\
        $\mu$ = FC($z$) & (1024,) \\
        $\sigma$ = FC($z$) & (1024,) \\
        $z' = \sigma \epsilon + \mu$ & (1024,)\\
        $z'' = FC(z')$ &  $(2 \cdot 2 \cdot 512)$ \\
        reshape: & (2, 2, 512)\\
        \midrule
        \multicolumn{2}{c}{decoder}\\
        kernel size & output filters \\
        \midrule
        (4, 4) &  (6, 6, 256)\\
        (4, 4) &  (14, 14, 128)\\
        (4, 4) & (30, 30, 64) \\
        (4, 4) &  (62, 62, 32)\\
        (4, 4) & (126, 126, 16) \\
        (4, 4) &  (254, 254, 8)\\
        (6, 6) &  (512, 512, 1)\\
        \bottomrule
    \end{tabular}%
    }\\[2ex]
    \resizebox{.3\linewidth}{!}{%
    \begin{tabular}[t]{ll}
        \multicolumn{2}{c}{DCGAN}\\
        \toprule
        \multicolumn{2}{c}{generator}\\
        kernel size & output filters \\
        \midrule
         (4, 4) & (4, 4, 1024) \\
         (4, 4) & (8, 8, 512) \\
         (4, 4) & (16, 16, 256) \\
         (4, 4) & (32, 32, 128) \\
         (4, 4) & (64, 64, 64) \\
         (4, 4) & (128, 128, 32) \\
         (4, 4) & (256, 256, 16) \\
         (4, 4) & (512, 512, 1) \\
        \midrule
        \multicolumn{2}{c}{discriminator}\\
        kernel size & output filters \\
        \midrule
         (4, 4) & (256, 256, 4) \\
         (4, 4) & (128, 128, 8) \\
         (4, 4) & (64, 64, 16) \\
         (4, 4) & (32, 32, 32) \\
         (4, 4) & (16, 16, 64) \\
         (4, 4) & (8, 8, 128) \\
         (4, 4) & (4, 4, 256) \\
         (4, 4) & (1, 1, 512) \\
         minibatch discrimination &  (1, 1, 528)\\
         FC &  (1,)\\
        \bottomrule
    \end{tabular}%
    }
    \resizebox{.3\linewidth}{!}{%
    \begin{tabular}[t]{ll}
      \multicolumn{2}{c}{BiGAN}\\
      \toprule
      \multicolumn{2}{c}{generator}\\
      kernel size & output filters \\
      \midrule
      (4, 4) &  (4, 4, 1024)\\
      (4, 4) & (8, 8, 512) \\
      (4, 4) &  (16, 16 ,256)\\
      (4, 4)${}^\dagger$ &  (32, 32, 128)\\
      (4, 4)${}^\dagger$ &  (64, 64, 64)\\
      (4, 4) &  (128, 128, 1)\\
      \midrule
      \multicolumn{2}{c}{encoder}\\
      kernel size & output filters \\
      \midrule
      (4, 4) &  (64, 64, 64)\\
      (4, 4) & (32, 32, 128) \\
      (4, 4) &  (16, 16 ,256)\\
      (4, 4)${}^\dagger$ &  (8, 8, 512)\\
      (4, 4)${}^\dagger$ &  (4, 4, 1024)\\
      (4, 4) &  (1, 1, 200)\\
      \midrule
      \multicolumn{2}{c}{Discriminator: Image Branch}\\
      kernel size & output filters \\
      \midrule
      (4, 4) &  (64, 64, 64)\\
      (4, 4) & (32, 32, 128) \\
      (4, 4) &  (16, 16 ,256)\\
      (4, 4)${}^\dagger$ &  (8, 8, 512)\\
      (4, 4)${}^\dagger$ &  (4, 4, 1024)\\
      (4, 4) &  (1, 1, 1024)\\
      \midrule
      \multicolumn{2}{c}{Discriminator: Code Branch}\\
      kernel size & output filters \\
      \midrule
        (1, 1) & (1, 1, 512) \\
        (1, 1) & (1, 1, 512) \\
      \midrule
      \multicolumn{2}{c}{Discriminator: Combination}\\
      kernel size & output filters \\
      \midrule
      \multicolumn{2}{c}{stack branches}  \\
        (1, 1) & (1, 1, 1024) \\
        (1, 1) & (1, 1, 1024) \\
        (1, 1) &  (1, 1, 1)\\
      \bottomrule
  \end{tabular}%
    }
    \resizebox{.3\linewidth}{!}{%
    \begin{tabular}[t]{ll}
        \multicolumn{2}{c}{$\alpha$-GAN}\\
        \toprule
      \multicolumn{2}{c}{generator}\\
      kernel size & output filters \\
      \midrule
      (4, 4) &  (4, 4, 1024)\\
      (4, 4) & (8, 8, 512) \\
      (4, 4) &  (16, 16 ,256)\\
      (4, 4)${}^\dagger$ &  (32, 32, 128)\\
      (4, 4)${}^\dagger$ &  (64, 64, 64)\\
      (4, 4) &  (128, 128, 1)\\
      \midrule
      \multicolumn{2}{c}{encoder}\\
      kernel size & output filters \\
      \midrule
      (4, 4) &  (64, 64, 64)\\
      (4, 4) & (32, 32, 128) \\
      (4, 4) &  (16, 16 ,256)\\
      (4, 4)${}^\dagger$ &  (8, 8, 512)\\
      (4, 4)${}^\dagger$ &  (4, 4, 1024)\\
      mean and variance \\
      (4, 4) &  (1, 1, 200)\\
      \midrule
      \multicolumn{2}{c}{discriminator}\\
      kernel size & output filters \\
      \midrule
      (4, 4) &  (64, 64, 64)\\
      (4, 4) & (32, 32, 128) \\
      (4, 4) &  (16, 16 ,256)\\
      (4, 4)${}^\dagger$ &  (8, 8, 512)\\
      (4, 4)${}^\dagger$ &  (4, 4, 1024)\\
      minibatch discrimination &  (4, 4, 1028)\\
      (4, 4) &  (1, 1, 1)\\
      \midrule
      \multicolumn{2}{c}{code-discriminator}\\
      kernel size & output filters \\
      \midrule
        (1, 1) & 100 \\
        (1, 1) & 50 \\
        (1, 1) & 25 \\
        (1, 1) & 1 \\
      \bottomrule
    \end{tabular}%
    }\\
    \begin{flushleft}
    Notes: * denotes additional max-pooling/nearest-neighbor upsampling, FC denotes fully-connected layer, $\epsilon \sim \mathcal{N}(\mathbf{0}, I_{512})$. ${}^\dagger$ denotes an additional self-attention layer.
    \end{flushleft}
}

\clearpage
\subsection{Data Augmentation}
We use two different augmentation strategies, named \emph{default} (used for GANs and VAE) and \emph{advanced} (used for CAE and BAE).

\subsubsection{Default}
\begin{itemize}
    \item Horizontally flip 50\% of all images
    \item Vertically flip 50\% of all images
    \item Center pad all images to the target resolution
\end{itemize}

\subsubsection{Advanced}
\begin{itemize}
    \item Horizontally flip 50\% of all images
    \item Vertically flip 50\% of all images
    \item For 50\% of the images change the brightness by multiplying all channels by a scalar value drawn randomly from the uniform distribution $U(0.8, 1.2)$
    \item For 50\% of the images randomly scale in x- and y- direction independently by a factor drawn randomly from the uniform distribution $U(0.8, 1.2)$
    \item For 50\% of the images rotate the image by an angle drawn randomly from the uniform distribution $U(-20, 20)$
    \item Center pad all images to the target resolution
\end{itemize}

\subsection{Training Details}
\begin{itemize}
    \item For all models we train 4 variants with different random seeds being 42, 4242, 424242, and 42424242.
\end{itemize}

\subsubsection{CAE}
\begin{itemize}
    \item Batch Size: 32
    \item Image Resolution: $512 \times 512$
    \item 1,000 epochs
    \item Batch Normalization
    \item Learning rate: 0.0001
    \item Adam optimizer
\end{itemize}

\subsubsection{BAE}
\begin{itemize}
    \item Batch Size: 32
    \item Image Resolution: $512 \times 512$
    \item 500 epochs
    \item Batch Normalization
    \item Learning rate: 0.0001
    \item Adam optimizer
\end{itemize}

\subsubsection{VAE}
\begin{itemize}
    \item Batch Size: 32
    \item Image Resolution: $512 \times 512$
    \item 500 epochs
    \item Batch Normalization
    \item $z_{dim} = 2,048$, $h_{dim}=18,432$
    \item Learning rate: 0.0001
    \item Adam optimizer
\end{itemize}

\subsubsection{DCGAN}
\begin{itemize}
    \item Batch Size: 80
    \item Image Resolution: $512 \times 512$
    \item 500 epochs
    \item No Batch Normalization
    \item Spectral Normalization
    \item Soft Labels
    \item Generator Learning Rate: 0.001
    \item Discriminator Learning Rate: 0.00001
    \item Soft Delta: 0.01
    \item $z_{dim} = 2,048$
    \item As we observed mode collapse, we added minibatch discrimination \cite{Salimans2016}
    \item Adam optimizer
\end{itemize}

\subsubsection{BiGAN}
\begin{itemize}
    \item Batch Size: 16
    \item Image Resolution: $128 \times 128$
    \item 500 epochs
    \item Generator \& Encoder Learning Rate: 0.001
    \item Discriminator Learning Rate: 0.000005
    \item Adversarial Loss: Hinge Loss
    \item $z_{dim} = 100$
    \item Adam optimizer
\end{itemize}

\subsubsection{$\alpha$-GAN}
\begin{itemize}
    \item Batch Size: 16
    \item Image Resolution: $128 \times 128$
    \item 500 epochs
    \item Generator \& Encoder Learning Rate: 0.001
    \item Discriminator \& Code-Discriminator Learning Rate: 0.000005
    \item Adversarial Loss: Hinge Loss
    \item $z_{dim} = 100$
    \item As we observed mode collapse, we added minibatch discrimination \cite{Salimans2016}
    \item Adam optimizer
\end{itemize} 

\end{document}